\documentclass[12pt]{appolb}
\usepackage{amssymb}
\date{ }
\title{ Trading by Quantum Rules - Quantum Anthropic Principle}

{\Huge \bf \author{E. W. Piotrowski\\ Institute of Theoretical
Physics, University of Bia\l ystok,\\ Lipowa 41, Pl 15424 Bia\l
ystok, Poland\\ e-mail: ep@alpha.uwb.edu.pl\\ and J. S\l adkowski
\\ Institute of Physics, University of Silesia, \\ Uniwersytecka
4, Pl 40007 Katowice, Poland \\ e-mail: sladk@us.edu.pl}}
\begin{document}
\maketitle
\def\Z{{\bf Z\!\!Z}}
\def\R{{\bf I\!R}}
\def\N{{\bf I\!N}}
\def\C{{\mathbb C}}
\begin{abstract}
This is a short review of the background and recent development in
quantum game theory and its possible application in economics and
finance. The intersection of science and society is discussed and
Quantum Anthropic Principle is put forward. The review is
addressed to non-specialists.

\end{abstract}
One hundred years ago, a single concept changed our world view
forever \cite{1}. Contemporary technology is based on
implementation of  quantum phenomena as a result of this seminal
idea. Now only social sciences persist in classical paradigm what
might be considered as an obstacle to unification of science in
the quantum domain. Quantum theory is up to now the only
scientific theory that requires of observer that he takes into
consideration the usually neglected influence of the method of
observation on the result of observation. Full and absolutely
objective information about the investigated phenomenon is
impossible and this is a fundamental principle of Nature and not
deficiency in our technology.  Critics of the possibility of
application of quantum methods in the domain of political and
social sciences often ignore this fact. A new way of perceiving
the reality implies fascinating conclusions \cite{2} that dim the
non-scientific visions of SF literature and Hollywood output. This
approach seems to be consistent with scientific experiments. If
fundamental phenomena (e.g. microworld) do not allow for classical
description why should economics phenomena  do? Maybe the modest
quantitative achievements  of social sciences if compared with
precise results of experiments performed by physicists result only
from the persistence in the classical paradigm? Will we manage  to
extend the domain of the quantum paradigm to include social
sciences? The
answer should be known within a few years. \\

At the beginning  of the XXI century first papers devoted to
quantum description of markets \cite{3}-\cite{6} were published.
But before we characterize this approach let us recall scientific
achievements that made it possible. We have got used to
simulations (modelling) of the surroundings and various phenomena
on computers. Contemporary computers act according the Turing
ideas but some 20 years ago Richard P. Feynman \cite{7} argued
that fundamental and well understood properties of Nature prevent
almost every physical process from being successfully simulated on
a Turing machine (i. e. computer). Such well known phenomena as
stability of matter, chemical reactions, conductivity, evolution
of stars and so on can only be understood on the quantum level.
Modern technologies are developed only due to investigation into
quantum nature of matter. The powerful computers that are at our
disposal, although classical in the sense to be explained below,
are constructed due to the achievements of quantum condensed
matter physics. So what is the problem? There is one:
computational complexity. Roughly speaking algorithms can become
so complicated that computation is in fact impossible \cite{8}.
(We do not want to go into the details of the beautiful
mathematical theory of computation but there are also problems
that cannot be solved at all!). Feynman drew the following
conclusion. Turing's ideas should be reformulated so that they
would incorporate the quantum character of Nature (according to
the modern physics all phenomena described in classical way are
only asymptotic or averaged results of quantum processes). For
example we have got an excellent theory of electro-magnetic
interactions (quantum electrodynamics) but to describe, say, the
scattering of two electrons at the precision required by
contemporary experiments several teams of experienced physicists
have to work for several years. This should be compared with the
actual behaviour of the electrons in question: they just scatter
in the twinkling of an eye without any computation (in the common
meaning of the word). There should be a quantum computer, whatever
it means, that performs better! In 1994 Peter Shor invented a fast
quantum algorithm for prime factoring of natural numbers \cite{9}.
This is an example of a problem  that is extremely difficult to
solve on a Turing machine. Due to its computational complexity
prime factoring is used in the nowadays most popular cryptographic
system (RSA). But, as Shor showed us, prime factoring is quite an
easy task on a quantum computer, provided we have got one. This
means that RSA codes can be broken! There are already first
successful probes of simple quantum computations \cite{10}. Of
course, these computation can be faster and cheaper done by hand
but nevertheless these experiment are promising. The thorough
investigation of quantum entangled states speeded up the
development of quantum cryptography that is immune even to attacks
by quantum algorithms. All these results induced scientist to
consider quantum strategies in games what gave birth to quantum
game theory that generalizes the classical von Neumann's ideas
\footnote{Paradoxically this outstanding mathematician was also
the author mathematical formalism of quantum theory} \cite{11}. A
lot of newspapers
announced this achievement in 1999. \\

Iqbal and Toor have applied the method of quantization of games
proposed by Marinatto and Weber \cite{12} in biology. Their
results are very interesting \cite{13}-\cite{15}. Recently they
have used the same formalism to analyse  the Stackelberg duopoly
(leader-follower model)\cite{16}. In the classical setting the
follower becomes worse-off compared to the leader who becomes
better-off. Iqbal and Toor have shown that in the quantum setting
the follower is not hurt even if he or she knows the action of the
leader. The backward induction outcome is the same as the Nash
equilibrium in the classical Cournot game that is when decision
are made simultaneously and there is no information hurting
players. \\

The above approach was  influenced by the development of quantum
cryptography so the contest takes place on a "quantum board" - the
set of possible states of the game. This might suggest that
non-classical aspects of various games could reveal themselves
only under very sophisticated conditions when ultra-modern quantum
technologies would make possible the existence of futuristic
markets or stock exchanges. On such would-be markets strategies,
being unitary operations, will be extremely sensitive to
perturbation destroying quantum coherence. This will assure
absolute privacy and unavoidable detection of any manipulation (no
deleting theorem). But this promising future, despite its subtle
technological constraint, must not necessarily be the only way of
benefiting from the rich and sometimes surprising opportunities
offered by quantum theory. We have managed to formulate a new
approach to quantum game theory that is suitable for description
of market transactions in term of supply and demand curves
\cite{3}-\cite{6}. In this approach quantum strategies are vectors
in some Hilbert space and can be interpreted as superpositions of
trading decisions. For an economist (or trader) they form the
potential "quantum board". Due to the possible economics context
the quantum strategies reveal a lot of interesting properties.
Supply strategies of market objects are Fourier transforms of
their respective demand states. Strategies and not the apparatus
or installation for actual playing  are at the very core of the
theory. If necessary the actual subject of investigation may
consist of single traders, teams of traders or event the whole
market. Of course,  sophisticated equipment built according to
quantum rules may be necessary for generating or clearing quantum
market but we must not exclude the possibility that human
consciousness (brain) performs that task equally well. Even more,
a sort of quantum playing board may be the natural theater of
"conflict games" played by our consciousness \cite{8}. We envisage
that in future, instead of penetration of the innermost recesses
of the brain, very interesting problems can be approached by
investigation of the possible quantum features of human behaviour.
It is possible that elementary "components" of consciousness are
formed by wave function spread over large domains and having no
concrete localization (as those of electrons forming electrical
current). It is worth to note here that recent investigations
reveal sort of randomness in brain's responses that may be of
quantum origin \cite{17}. If human strategies are collective
properties of molecules forming neural system then quantum
automata might be the only tools to describe real social games.
This should be compared with phonons (collective excitations in
solids) that escapes our perception if the cristalline network is
decomposed into its basic ingredients. We should stress here that
quantization does not simple mean introducing elements of
randomness into the model. Quantum theory is different from
statistical description on both qualitative and quantitative level \cite{2,7,8}. \\

In the newly proposed approach  spontaneous or institutionalized
market transactions are described in terms of projective operation
acting on Hilbert spaces of strategies of the traders. Quantum
entanglement is necessary to strike the balance of trade. The
text-book examples of departures from the demand-supply law are
related to the negative probabilities that often emerge in quantum
theories and form very interesting illustrations of them
\cite{4,5}. This theory predicts the property of undividity of
attention of traders (no clonning theorem). The sudden and violent
changes of prices can be explained by the quantum Zeno phenomenon.
The theory unifies also the English auction with the Vickrey's one
attenuating the motivation properties of the latter. There are
apparent analogies with quantum thermodynamics that allow to
interpret market equilibrium  as a state with vanishing financial
risk flow. Sometimes euphoria, panic or herd instinct cause
violent changes of market prices. Such phenomena can be described
by non-commutative quantum mechanics. There is a simple tactics
that maximize the trader's profit on an effective market. It can
be expressed as {\it accept profits equal or greater then the one
you have formerly achieved on average}. \\

Even if at early  civilization stages markets are governed by
classical laws (this may be questioned \cite{8}) the incomparable
efficacy of quantum algorithms in multiplying profits should
result in such market evolution so that  quantum behaviour will be
prevailing over the classical one. This {\it quantum anthropic
principle} \cite{3} could have been observed at work in the former
century:  quantum description was more effective and economical
from both  technological and economic points of view and the
quantum paradigm replaced the classical one. Nowadays an essential
part of transactions made on NYSE or NASDAQ are in fact made by
computers.  We envisage that these computers will be replaced by
quantum ones. Quantum market games broaden our horizons and offer
new opportunities for the economy. On the other hand, "{\it It
might be that while observing the due ceremonial of everyday
market transaction we are in fact observing capital flows
resulting from quantum games eluding classical description. If
human decisions can be traced to microscopic quantum events one
would expect that nature would have taken advantage of quantum
computation in evolving complex brains. In that sense one could
indeed say that quantum computers are playing their market games
according to quantum rules}" \cite{18}. David Deutsch has proposed
an interesting unification of theories of information, evolution
and quanta \cite{2}. But why quantum social sciences should emerge
just now? They could have not emerged earlier because a tournament
quantum computer versus classical one is not possible without
technological development necessary for a construction of quantum
computers. Quantum-like approach to market description might turn
out to be an important theoretical tool for investigation of
computability problems in economics or game theory even  if never
implemented in real market \cite{19}.

We encourage the reader to visit the web site
http://alpha.uwb.edu.pl/ep/sj/index.shtml where she or he can find
full texts of our papers and links to other related papers and
sites.

\end{document}